# Search for a signal on intermediate baryon systems formation in hadron-nuclear and nuclear-nuclear interactions at high energies

Y H Huseynaliyev[1], M K Suleymanov[1], E U Khan[1], A Kravchakova[2] and S Vokal[2,3]

[1] COMSATS Institute of Information Technology, Islamabad, Pakistan
[2] University of P. J. Shafarik, Koshice, Slovak Republic
[3] Veksler and Baldin Laboratory of High-Energies, JINR, Dubna, Russia

E-mail: yashar_huseynaliyev@comsats.edu.pk

**Abstract**. We have analyzed the behavior of different characteristics of hadron-nuclear and nuclear-nuclear interactions as a function of centrality to get a signal on the formation of intermediate baryon systems. We observed that the data demonstrate the regime change and saturation. The angular distributions of slow particles exhibit some structure in the above mentioned reactions at low energy. We believe that the structure could be connected with the formation and decay of the percolation cluster. With increasing the mass of colliding nuclei, the structure starts to become weak and almost disappears ultimately. This shows that the number of secondary internuclear interactions increases with increasing the mass of the colliding nuclei. The latter could be a reason of the disintegration of any intermediate formations as well as clusters, which decrease their influence on the angular distribution of the emitted particles.

## 1. Introduction

The characteristics of secondary particles in the hadron-nuclear and nuclear-nuclear interactions depend upon the nature of intermediate formation and their study leads to the understanding of the dynamics of the interactions. There are many papers which predict that the angular distributions of fragments could have some special structure as a result of the formation and decay of some intermediate formations (for example see [1]). There are only a few experimental results [2-5] that could be considered as some confirmation of it. These results mainly belong to the interactions of hadrons and light nuclei. We have used the experimental data of Kr+Em - reaction at 0.95 A GeV [6] and Au+Em - reaction at 10.7 A GeV [7] to analyze angular distribution of secondary slow particles as a function of centrality [8].

## 2. Experimental results and discussion

The results were shown separately for the central collisions and for the peripheral ones. To select the central collisions, we used the criteria $N_F \geq 20$ suggested by Abdinov et al [9] showed that the central Au+Em-events (at 10.7 A GeV) should be selected using the criteria $N_F \geq 40$. We followed a different criterion because the number of events is very small with $N_F \geq 40$.

We could see that in peripheral collisions the angular structure becomes cleaner. It could mean that these structures are due to nucleon elastic scattering. It was seen that the structure got weak and almost disappeared in central collisions.

The angular distributions of the particles emitted in Au+Em-reactions at 10.7 A GeV energy are shown in figure. Black circles are experimental data for b –particles (mostly protons with $p \leq 0.2$ GeV/c and multiple charged target fragments having a range, 3mm) emitted in the (a) central collisions and (b) peripheral collisions. The histograms are the results coming from the cascade evaporation model [10]. It is seen that the angular distributions of these particles emitted in Au+Em-reactions (as well as in the Kr+Em-reactions) do not contain any special structure except the one that is described using the usual mechanisms of the interaction (e.g. cascade evaporation).

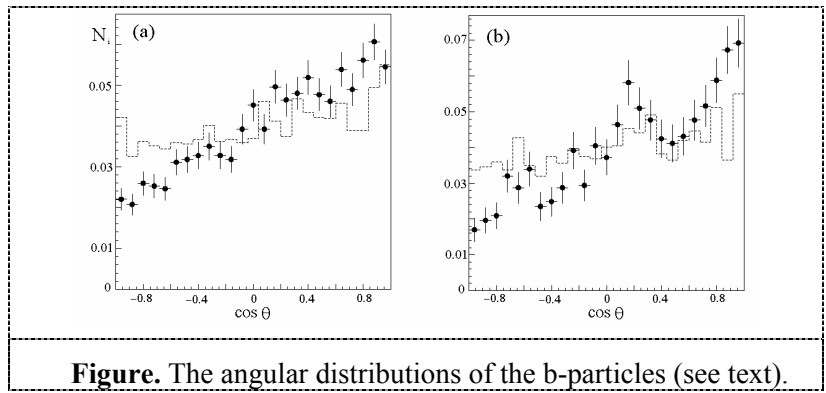

**Figure.** The angular distributions of the b-particles (see text).

If we compare the results of the low mass nuclei reaction mentioned above with those mentioned here we see that with increasing the mass of the projectile the angular distribution of slow particles change and the structure which was demonstrated in the case of lighter projectile almost disappear. This could be explained as follows: during the interaction of the heavier projectile with nuclear target, number of secondary interactions as well as number of nucleon-nucleon elastic scatterings and re-scattering events increases. These effects could lead to the information about the disintegration of any intermediate formations as well as clusters, decreasing their influence on the angular distributions of the emitted particles.

Hence to get more clear information about the intermediate nuclear systems we need to use the data of hadron-nuclear and light nuclear-nuclear interactions.